\documentclass[12pt]{article}
\usepackage[utf8]{inputenc}

\usepackage{graphicx} % Required for inserting images
\usepackage{natbib}
\usepackage{amssymb}
\usepackage{amsmath}
\usepackage{amsfonts}
\usepackage{booktabs}
\usepackage{xcolor}
\usepackage{float}
\usepackage{subfig}
\usepackage{graphics}
\usepackage{graphicx}
\usepackage{enumitem}
\usepackage{hyperref}

\title{Cubature scheme for spatio-temporal Poisson point processes estimation}
\author{Nicoletta D'Angelo \and Giada Adelfio}
\date{\small Department of Economics, Business and Statistics, University of Palermo}

\begin{document}

\maketitle

\begin{abstract}
    This work presents the cubature scheme for the fitting of spatio-temporal Poisson point processes.
    The methodology is implemented in the \cite{R} package \textbf{stopp} \citep{stopp}, published on the Comprehensive R Archive Network (CRAN) and available from \url{https://CRAN.R-project.org/package=stopp}.
Since the number of dummy points should be sufficient for an accurate estimate of the likelihood, numerical experiments are currently under development to give guidelines on this aspect.
\end{abstract}
 
\section{Introduction}

This work presents the spatio-temporal counterpart of \cite{berman1992approximating}'s algorithm for fitting spatio-temporal Poisson point process models of very general form.
The generalization of this algorithm or, equivalently of \cite{baddely:turner:00}'s one, to the spatio-temporal context represents the first novel possibility to fit spatio-temporal Poisson point process models, including both homogeneous and inhomogeneous processes, possibly depending on either parametric and non-parametric specifications of both coordinates and external covariates.
To do so, we use a finite cubature approximation to the log-likelihood, conceptually extending the quadrature scheme implemented in the \texttt{spatstat} package \citep{spat}.
Note that the current proposal treats each dimension equivalently so that the procedure can be straightforwardly employed for fitting purely spatial three-dimensional point processes. However, this research only mentions spatio-temporal point process theory. 

The structure of the work is as follows.
Section \ref{sec:stpp} gives preliminaries on spatio-temporal point processes. Section \ref{sec:quad} contains the proposal of the cubature scheme for the estimation of a general spatio-temporal Poisson point process model. Section \ref{sec:covs} illustrates the procedure proposed to overcome the additional challenges of including external covariates in the linear predictor of the fitted Poisson model. Section \ref{sec:mark} presents the multitype spatio-temporal Poisson point process.
Finally, a computational note is in Section \ref{sec:comp}.

\section{Spatio-temporal point processes}
\label{sec:stpp}

We consider a spatio-temporal point process with no multiple points as a random countable subset $X$ of $\mathbb{R}^2 \times \mathbb{R}$, where a point $({u}, t) \in X$ corresponds to an event at $ {u} \in \mathbb{R}^2$ occurring at time $t \in \mathbb{R}$.
A typical realisation of a spatio-temporal point process $X$ on $\mathbb{R}^2 \times \mathbb{R}$ is a finite set $\{({u}_i, t_i)\}^n_{
i=1}$ of distinct points within a
bounded spatio-temporal region $W \times T \subset \mathbb{R}^2 \times \mathbb{R}$, with area $\vert W\vert  > 0$ and length $\vert T\vert  > 0$, where $n \geq 0$ is not fixed in
advance. The point patter will be denoted as $\mathbf{x}=\{({u}_1, t_1),\ldots, ({u}_n, t_n)\}$.
In this context, $N(A \times B)$ denotes the number of points of a set $(A \times B) \cap X$, where $A \subseteq W$ and $B \subseteq T$. As usual \citep{daley:vere-jones:08}, when $N(W \times T) < \infty $ with probability 1, which holds e.g. if $X$ is defined on a bounded set, we call $X$ a finite spatio-temporal point process.

For a given event $({u}, t)$, the events that are close to $({u}, t)$ in both space and time, for each spatial distance $r$ and time lag $h$, are given by the corresponding spatio-temporal cylindrical neighbourhood of the event $({u}, t)$, which can be expressed by the Cartesian product as
$$
b(({u}, t), r, h) = \{({v}, s) : \vert \vert{u} - {v}\vert \vert \leq r, \vert t - s \vert \leq h\} , \quad \quad
({u}, t), ({v}, s) \in W \times T,
$$
where $ \vert \vert \cdot \vert \vert$ denotes the Euclidean distance in $\mathbb{R}^2$. Note that $b(({u}, t), r, h)$ is a cylinder with centre ({u}, t), radius $r$, and height $2h$.

Product densities $\lambda^{(k)}, k  \in \mathbb{N} \text{ and }  k  \geq 1 $, arguably the main tools in the statistical analysis of point processes, may be defined through the so-called Campbell Theorem (see \cite{daley:vere-jones:08}),  that constitutes an essential result in spatio-temporal point process theory, stating that, given a spatio-temporal point process $X$, for any non-negative function $f$ on $( \mathbb{R}^2 \times \mathbb{R} )^k$
\begin{equation*}
  \mathbb{E} \Bigg[ \sum_{\zeta_1,\dots,\zeta_k \in X}^{\ne} f( \zeta_1,\dots,\zeta_k)\Bigg]=\int_{\mathbb{R}^2 \times \mathbb{R}} \dots \int_{\mathbb{R}^2 \times \mathbb{R}} f(\zeta_1,\dots,\zeta_k) \lambda^{(k)} (\zeta_1,\dots,\zeta_k) \prod_{i=1}^{k}\text{d}\zeta_i,
\label{eq:campbell0}  
\end{equation*}
where $\neq$ indicates that the sum is over distinct values. In particular, for $k=1$ and $k=2$, these functions are respectively called the \textit{intensity function} $\lambda$ and the \textit{(second-order) product density} $\lambda^{(2)}$.

Broadly speaking, the intensity function describes the rate at which the events occur in the given spatio-temporal region, while the second-order product densities are used for describing spatio-temporal variability and correlations between pairs of points of a pattern. They represent the point process analogues of the mean function and the covariance function of a real-valued process, respectively.

The first-order intensity function is defined as 
\begin{equation*}
 \lambda({u},t)=\lim_{\vert \text{d}{u} \times \text{d}t\vert  \rightarrow 0} \frac{\mathbb{E}[N(\text{d}{u} \times \text{d}t )]}{\vert \text{d}{u} \times \text{d}t\vert }, 
\end{equation*}
where $\text{d}{u} \times \text{d}t $ defines a small region around the point $({u},t)$ and $\vert \text{d}{u} \times \text{d}t\vert $ is its volume. 

\subsection{Marked spatio-temporal point processes}

The marks of a spatio-temporal point process are defined as random measurements $m_i$ in
a set $\mathcal{M}$, associated to locations of events $x_i = (u_i ,t_i)$ in a studied area $W \times T \subset \mathbb{R}^2 \times \mathbb{R}$.

A marked spatio-temporal point process is composed of a point process and
associate marks, which can be expressed as
$\{(u_i ,t_i, m_i ) : i = 1,\ldots ,n\},$
where $(u_1 ,t_1) ,\ldots ,(u_n ,t_n)$ are locations and $m_1 ,\ldots ,m_n$ are associated
marks.
A marked point process $Y$ can be understood as a pure point
process on $W \times T \times \mathcal{M}$, where $\mathcal{M}$ is the
domain of marks, such that $N(W \times T \times \mathcal{M}) < \infty,$ if $W \times T$ is bounded.

One can define the first-order intensity function as
$$\lambda(u,t,m) = \lim_{|du \times dt \times dm| \rightarrow 0} \frac{\mathbb{E}[N(du \times dt \times dm)]}{|d \textbf{s} \times dm|}.$$

Given this general setup, one may obtain various forms of marked point processes, most notably multivariate/multitype point processes with  $\mathcal{M}=\{1,\ldots,m\}$ \citep{diggle:13}.

\section{First-order intensity function estimation}
\label{sec:quad}

We assume that the template model is a Poisson process, with a parametric intensity or rate function $$\lambda(u, t; \boldsymbol{\theta}), \qquad  u \in
W,\quad  t \in T, \quad \boldsymbol{\theta} \in \Theta,$$
with $\boldsymbol{\theta}$ some unknown parameters.

The log-likelihood is 
\begin{equation}
    \log L(\boldsymbol{\theta}) = \sum_i
\lambda({u}_i, t_i; \boldsymbol{\theta}) - \int_W\int_T
\lambda({u}, t; \boldsymbol{\theta}) \text{d}t\text{d}u
\label{eq:glo_lik}
\end{equation}
up to an additive constant, where the sum is over all points $({u}_i, t_i)$  in the point pattern $\textbf{x}$  \citep{daley:vere-jones:08}. 

We often consider intensity models of log-linear form
\begin{equation*}
   \lambda({u}, t; \boldsymbol{\theta}) = \exp(\boldsymbol{\theta}^{\top} \textbf{Z}({u}, t)), \quad
{u} \in W,\quad  t \in T
\end{equation*}
where $\textbf{Z}(u,t)=\{Z_1(u,t), \ldots, Z_p(u,t)\}$ are known covariate functions. 

\subsection{Cubature scheme}

We use a finite \textit{cubature approximation} to the log-likelihood. Renaming the data points as $x_1,\dots , x_n$ with $({u}_i,t_i) = x_i$ for $i = 1, \dots , n$, then generate $m$  additional ‘‘dummy points’’ $({u}_{n+1},t_{n+1}) \dots , ({u}_{m+n},t_{m+n})$ to
form a set of $n + m$ cubature points (where $m > n$). Then we determine cubature weights $a_1, \dots , a_m$
so that integrals in \eqref{eq:glo_lik} can be approximated by a Riemann sum
\begin{equation*}
    \int_W \int_T \lambda({u},t;\boldsymbol{\theta})\text{d}t\text{d}u \approx \sum_{k = 1}^{n + m}a_k\lambda({u}_{k},t_{k};\boldsymbol{\theta})
\end{equation*}
where $a_k$ are the cubature weights such that $\sum_{k = 1}^{n + m}a_k = l(W \times T)$ where $l$ is the Lebesgue measure.

The log-likelihood \eqref{eq:glo_lik} of the template model can be approximated by
\begin{equation}
\begin{split}
        \log L(\boldsymbol{\theta})   \approx &
\sum_k
\log \lambda({u}_k,t_k; \boldsymbol{\theta}) +
\sum_k
(1 - \lambda({u}_k,t_k; \boldsymbol{\theta}))a_k
 = \\
		 &
\sum_k
e_k \log \lambda({u}_k, t_k; \boldsymbol{\theta}) + (1 - \lambda({u}_k, t_k; \boldsymbol{\theta}))a_k
\end{split}
\label{eq:approx0}
\end{equation}

where $e_k$ is the indicator that equals $1$ if $(u_k,t_k)$ is a data point.

Writing $y_k = e_k/a_k$ \eqref{eq:approx0} becomes
\begin{equation*}
    \log L(\boldsymbol{\theta}) \approx
\sum_k
a_k
(y_k \log \lambda({u}_k, t_k; \boldsymbol{\theta}) - \lambda({u}_k, t_k; \boldsymbol{\theta}))
+
\sum_k
a_k.
\end{equation*}

Apart from the constant $\sum_k a_k$, this expression is formally equivalent to the weighted log-likelihood of
a Poisson regression model with responses $y_k$ and and weights $a_k$. This can be
maximised using standard GLM software.
Another option not detailed in this work is the spatio-temporal extension of logistic spatial regression \citep{baddeley2014logistic}. 

To fit such models,  the spatio-temporal cubature scheme is obtained in the same manner, that is, by defining a spatio-temporal partition of $W \times T$ into cubes $C_k$ of equal volume $\nu$, assigning the weight $a_k=\nu/n_k$  to each cubature point (dummy or data), where $n_k$ is the number of points that lie in the same cube as the point $u_k$. 
The number of dummy points should be sufficient for an accurate estimate of the likelihood, and numerical experiments are currently under development to give guidelines on this aspect.

\section{Dependence on external covariates}\label{sec:covs}

% We now assume that the template model is a Poisson process, with a parametric intensity or rate function $$ \lambda({u}, t; \boldsymbol{\theta}) = \exp(\boldsymbol{\theta}^T Z({u}, t))$$ with spatial and temporal coordinates ${u} \in W, t \in T$, unknown parameters $\boldsymbol{\theta} \in \Theta,$ and some spatio-temporal covariates $Z({u}, t)$.

Given the above, it is easy to understand that to fit spatio-temporal Poisson point process models with first-order intensity depending on external covariates, their values must be known in every data and dummy point location. As this is typically unfeasible in practice, we propose first to interpolate covariate values at a very fine regular grid and then attribute to each data or dummy point the value of the closest point in three-dimensions. 
Hence, the interpolation at a (data or dummy) point location $x_k = (u_k, t_k)$ is performed through a \textit{three-dimensional inverse-distance weighting smoothing procedure} of the covariate values $Z(x_j)$ assumed to be observed at its sampling locations $j=1, \ldots, J$. In such a case, the smoothed value at a  data or dummy  location $x_k$ is 
$$
Z(x_k) = \frac{\sum_j w_j Z(x_j)}{\sum_j w_j},
$$
where the weight $w_j$ is the $j$-th element of the inverse $p$th powers of distance,
$$
\textbf{w}=\{w_j\}_{j=1}^J=\big\{\frac{1}{d||x_k-x_j||^p}\big\}_{j=1}^J
.$$

\section{Multitype spatio-temporal Poisson point processes}\label{sec:mark}

There are no additional theoretical difficulties in defining the intensity of a spatio-temporal multitype point process compared to the purely spatial one.
We simply consider the intensity of each type of point separately. Adopting the multivariate viewpoint, we split a multitype point process $Y$ into sub-processes $X_1, \ldots ,X_M$ where $X_M$ is the point process of points of type $m$ for each mark level $m = 1,2, \ldots ,M$.
We also write $X_g$ for the ground/unmarked point process, which only consists of the locations. In other words, $X_g$ is the superposition of all the processes $X_m$ for $m = 1,2, \ldots ,M$.

If the multitype point process $Y$ has intensity function $\lambda(u,t,m)$ for locations $(u,t)$ and marks $m$, then the unmarked process $X_g$ is also first-order stationary, with (marginal) intensity $$\lambda(u,t) = \sum_{m=1}^{M}\lambda(u,t,m) $$ where $\lambda(u,t,m)$ is the intensity function of each sub-process $X_m$.

\subsection{Replicated cubature scheme}

Based on the proposed cubature rule, we create a replicated cubature scheme.
Let  $M$  be the number of types of a categorical mark, each generating the point patterns $\textbf{x}_1,\dots,\textbf{x}_M$. We replicate the dummy points for each possible mark level $m$, and generate dummy marked points at the same locations as the data points but with different marks. Finally, the two dummy patterns and the data points are superimposed, and we compute the cubature weights and form the indicators 
$e_{mk}$ assuming value $1$ if $(u_{mk}, t_{mk}) \in {\textbf{x}_{m1}, \dots , \textbf{x}_{m n(\textbf{x}_m)}}$, that is, if $(u_{mk}, t_{mk})$ is a data point belonging to the mark level $m$. The indicator $e_{mk}$ will assume value $0$ in all the other cases $(u_{mk}, t_{mk})\notin {\textbf{x}_{m1}, \dots , \textbf{x}_{m n(\textbf{x}_m)}}$, that is, when $(u_{mk}, t_{mk})$ is either a dummy points or a data point not belonging to that specific mark level $m$.

Then, the log-likelihood of independent weighted Poisson variables becomes
 \begin{equation*}
    \log L(\boldsymbol{\theta}) \approx
\sum_m \sum_k
a_{mk}
(y_{mk} \log \lambda({u}_{mk}, t_{mk}; \boldsymbol{\theta}) - \lambda({u}_{mk}, t_{mk}; \boldsymbol{\theta}))
+
\sum_m \sum_k
a_{mk}.
\end{equation*}
		where $a_{mk}$ are the cubature weights obtained as $y_{mk}=\frac{e_{mk}}{a_{mk}}.$

The model should be fitted through a Generalized Linear Mixed Model \citep{breslow1993approximate}, in which the linear predictor contains random effects in addition to the usual fixed effects.
In our case, response variable values are still represented by $y_{mk}$, with weights $a_{mk}$, but with an additional group variable given by the mark.
Therefore, the differences among the mark levels can be modelled as random effects.

\section{Computational note}\label{sec:comp}

In  \textbf{stopp} \citep{stopp}, we make usage of the \texttt{gam} function of the \textbf{mgcv} package \citep{wood2017generalized} since it provides the possibility of including both smooth terms of the covariates (typical in point process theory for the spatio-temporal coordinates) and random effects. The latter comes in aid when wishing to fit a multitype point pattern, where basically, each type of the categorical mark will have its own set of fitted parameters.

\section*{Funding}
This work was supported by the Targeted Research Funds 2024 (FFR 2024) of the University of Palermo (Italy), the Mobilità e Formazione Internazionali - Miur INT project ``Sviluppo di metodologie per processi di punto spazio-temporali marcati funzionali per la previsione probabilistica dei terremoti", and the European Union -  NextGenerationEU, in the framework of the GRINS -Growing Resilient, INclusive and Sustainable project (GRINS PE00000018 – CUP  C93C22005270001).
The views and opinions expressed are solely those of the authors and do not necessarily reflect those of the European Union, nor can the European Union be held responsible for them.

\bibliography{BBB}

\end{document}